\documentclass[twocolumn,final,10pt]{article}
\usepackage{times}
\usepackage{cite}
\usepackage{graphics}
\usepackage{graphicx}

\title{It's Not What You Have, But How You Use It: Compromises in Mobile Device Use}

\author{
  Manas Tungare and Manuel P\'erez-Qui\~nones
  \\
  Center for Human-Computer Interaction and Dept. of Computer Science, Virginia Tech
  \\
  \{manas, perez\}@vt.edu
}

\begin{document}

\maketitle

\abstract{As users begin to use many more devices for personal information management (PIM) than just the traditional desktop computer, it is essential for HCI researchers to understand how these devices are being used in the wild and their roles in users' information environments. We conducted a study of 220 knowledge workers about their devices, the activities they performed on each, and the groups of devices used together. Our findings indicate that several devices are often used in groups; integrated multi-function portable devices have begun to replace single-function devices for communication (e.g. email and IM). Users use certain features opportunistically because they happen to be carrying a multi-function device with them. The use of multiple devices and multi-function devices is fraught with compromises as users must choose and make trade-offs among various factors.}

\section{Introduction}

The past few years have seen an increase in the usage of mobile devices for information management tasks. Although traditional desktops have continued to be a mainstay of users' information environments, other devices play an increasingly important role. As more and more devices are brought into a user's personal information ecosystem \cite{tungare_2006_personal}, it is important to understand the role each device plays in it, individually, as well as in relation to the other devices present.

Users tend to use their devices in certain commonly-occuring configurations: e.g. at the office, a user might use her desktop computer and personal digital assistant (PDA); on the road or at conferences, a researcher may use a laptop and a cellphone; etc. Some combinations and groups may be mandated by the manufacturer, and in fact, one may not function without the other. The iPod portable music player and iTunes music library software are an example of this sort of mandatory pairing. Studying the ad hoc formation of these groups can help designers of future products  design for a group of devices together rather than for each device individually \cite{pyla_2006_multiple}.
 
In this paper, we report some of our findings from a survey of 220 knowledge workers about their device usage habits and practices. Almost everyone used more than one device (at least a computer and a cellphone), and several participants used three or more devices. We asked them about the activities they commonly performed on each of their devices, and the groups of devices that they tended to use together. A lot of users reported having to make compromises between various factors when choosing a device or a combination of devices to work with.

\section{Background}

Our work lies at the intersection of personal information management (PIM) and multi-device (or multi-platform) user interfaces and covers an area that has not been studied by either community.

\subsection{Personal Information Management}

Recent research in Personal Information Management (PIM) has identified the problem of information fragmentation \cite{bergman_2006_the-project}, but the original definition only included the fragmentation of information across different collections, e.g. files, email messages, and bookmarks all seemed to be managed within similar, yet duplicate, hierarchies \cite{boardman_2003_too-many}. The issue of information fragmentation across multiple devices has been reported \cite{karger_2006_data} and preliminary suggestions to solve it have been provided, but this problem has not yet been studied in depth to ascertain exactly how and where the fragments of this information live.

PIM is a significant problem even when only a single device is in the picture. When multiple devices enter the landscape, managing information across them is a greater challenge.

\subsection{Multi-Device (or Multi-Platform) User Interfaces}

Most of the research in multi-platform user interfaces has focused on the specifics of the interaction on each device, on moving tasks between devices seamlessly, and maintaining some form of consistency when migrating such tasks. These user interfaces have been variously called \textit{plastic user interfaces} \cite{thevenin_1999_plasticity}, \textit{nomadic applications} \cite{mori_2003_tool}, or \textit{multi-browsing interfaces} \cite{johanson_2001_multibrowsing}. Various techniques for interface migration have been proposed, including model-based approaches \cite{mori_2003_tool, einsenstein_2001_applying}, user interface markup languages \cite{abrams_1999_uiml}, and transformation-based approaches \cite{richter_2005_transformation, florins_2004_graceful}.

Thus, while there is a significant body of knowledge on building multi-device interfaces, not much attention has been given to the information needs of users on these devices. Researchers have followed a task-oriented approach rather than an in\-for\-ma\-tion-oriented path.

\section{Motivation}

The questions we hoped to find answers to were mainly about the usage of multiple devices together. From our informal data gathering prior to the structured study, we realized that the presence of certain devices significantly altered the information management practices of users. We wanted to study why certain activities on certain types of information are only performed on a smaller subset of devices, even though it may be advantageous to perform them on many more devices. E.g., why do users who often use a calendar program on their desktop not use it on their cell phones?

We hypothesize that there exist strong inter-relationships between devices, and they lead users to make compromises in their mobile device use. Information flows are altered when a new device is introduced into a user's environment; equilibrium is then achieved after a certain period of time.

To provide the background for this study, we start with a description of the study, the participants, and their demographic composition.

\section{Study Description}

The study was conducted via a questionnaire survey distributed widely via employee email lists at several information technology companies, and via campus notices in multiple departments in a university. Since we wanted to study the usage patterns of users who used multiple devices to varying degrees, we concentrated on the population that was most likely to use many such devices in their everyday life: our audience largely consisted of knowledge workers, including professionals, students, professors, and administrative personnel. Since the survey was administered via the Internet, we were able to reach beyond just the local population, and received responses from all across the USA, and from a few other countries including the UK and India.

\subsection{Participant Demographics}

\begin{figure}[htb] 
   \centering
   \includegraphics[width=0.48\textwidth]{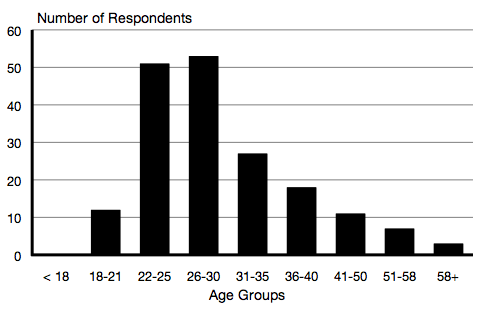} 
   \caption{Number of respondents by age group.}
   \label{figure:age}
\end{figure}

220 respondents completed the survey; 53\% of respondents were male, 30\% were female, and 17\% indicated neither.
157 respondents, or 71.3\%, reported that they considered themselves either full-time or part-time knowledge workers.
The study spanned an age range from 18 years to over 58 years old. Though a majority of the respondents were between 22 and 30 years old, other age groups were adequately represented (see Figure \ref{figure:age}). 

The participant pool consisted of users of varying levels of education completed, from high school to doctoral degrees. Due to our focus on knowledge workers, our study elicited a high number of responses from people who had completed advanced graduate degrees: Masters 34\% and Ph.D. 10\%.

\vspace{0.125in}

\section{Findings}

During our analysis, we focused on the use of multiple devices, if the use of multi-function devices affected the use of any other devices they carried, groups of devices that tended to be used together, the activities users performed on each device, and the compromises they made in all these aspects.

\subsection{Devices Used}

\begin{figure}[htb] 
   \centering
   \includegraphics[width=0.47\textwidth]{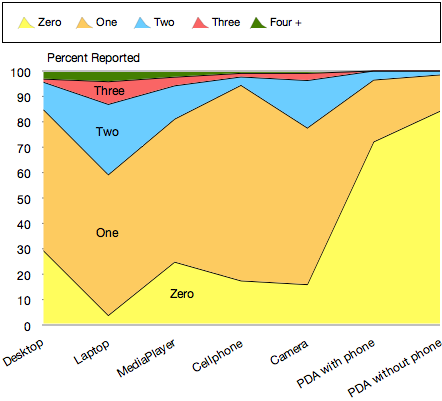} 
   \caption{Number of devices in each category reported as percentages.}
   \label{figure:devices-percent-cumulative}
\end{figure}

Figure \ref{figure:devices-percent-cumulative} shows the number of each type of device reported, converted to percentages. Our study found more laptop users than desktop users. Over 71\% of respondents used at least one desktop, while about 96\% used at least one laptop, which is higher than even the number of cell phones reported. This is representative of the current trend towards mobility and away from stationary platforms such as desktops. Portable media players have made their way into the hands of more than 80\%, almost equal to that of digital cameras.

Handheld computing devices that combine a Personal Digital Assistant (PDA) and a cell phone, such as the Blackberry, Palm Treo, Apple iPhone and others, are used by a minority of users, about 22\%. PDAs without built-in cell phone technology are used by fewer users, about 15\%.

\subsection{The Impact of Multi-Function Devices}

Those who use multi-function devices use them extensively for PIM tasks such as email, calendaring and IM, and also for news and (limited) Web browsing. Some participants reported that the presence of these handhelds had caused them to leave their laptops behind when they did not expect to work on complex documents (e.g. when on vacation), but many others reported that they still carried their laptops with them as the tool of choice for more complex computing activities.

\begin{quotation}
\textit{``Treo allowed me to stop carrying a separate pager. I still carry a laptop around.    However, when I don't have the laptop, I can still do almost everything -- except edit documents -- on my Treo.''}
\end{quotation}

These multi-function devices often replaced other sin\-gle-func\-tion devices, like cell phones, music players, and compact digital cameras. Participants reported that they started using more features of their device because they carried it with them for another purpose (we term this \textit{opportunistic use}). Certain activities also were moved from one device to another simply because it was now possible to do so, without the burden of carrying yet another device. This is an example of an unforeseen (or unintentional) advantage of acquiring a new device.

\begin{quotation}
\textit{``I previously owned an iPod and PDA, but never used them because I didn't have enough pockets to keep all my gadgets. With a multipurpose gadget, I now use those features because I actually have them with me all the time.''}
\end{quotation}

\begin{quotation}
\textit{``[After buying an iPhone,] music playing on [the] laptop has been drastically reduced. My portable music player [has been] completely replaced.''}
\end{quotation}

The quality of individual functional components of an integrated device was often compared to that of stand-alone devices and generally found to be lacking. Despite that, the convenience of carrying a smaller device led users to prefer them on certain occasions.

\begin{quotation}
\textit{``The iPhone camera comes nowhere near my Sony Alpha 100 DSLR, but I have it with me all the time.''}
\end{quotation}

Features of devices that did not integrate well within their existing information infrastructure were used less often. Users reported that synchronizing with their other devices was an important requirement, irrespective of the quality of the stand-alone feature.

\begin{quotation}
\textit{``I have a Windows Mobile Smartphone with a full keyboard. [...] Its camera isn't near as good enough to replace my digital camera and the calendar doesn't sync with my MacBook, so I don't use it.''}
\end{quotation}

\subsection{Groups of Devices}

\begin{figure}[htb]
\begin{center}
\includegraphics[width=0.5\textwidth]{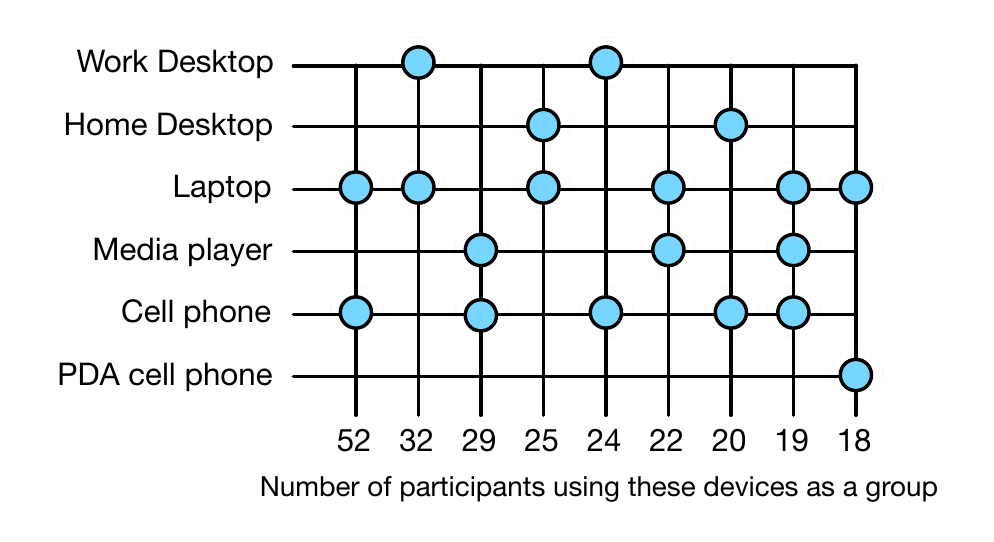}
\caption{Devices used in groups. Note: groups reported by fewer than 10 users are not included in this figure.}
\label{figure:device-groups}
\end{center}
\end{figure}

Several users reported that they used devices together in groups. Figure \ref{figure:device-groups} shows the most common device groups. Laptops and cell phones were used together by the most number of users, almost 24\%. The laptop and the cell phone also appeared the most times in combination with other devices. The low use of PDAs without an integrated cell phone for almost all tasks (as compared to the use of cell phones and PDAs with cell phones) indicates that these devices are considered less popular.

However, a lot of users were dissatisfied with the currently available synchronization tools for multiple devices. The high use of the laptop is indicative of the trend to keep all data on a single device to escape the need for synchronizing. Similarly, address books on cell phones were kept separate from those on laptops (or desktops). The same data (or application) was used for two distinct tasks (sending email from the laptop versus making a phone call using a cell phone), and therefore some users preferred to keep the two contact databases separate, again a compromise.

\begin{quotation}
\textit{
``Usually my contacts on the phone are just with numbers while my contacts on the computer are just with email addresses (makes sense since I'm using the former to make calls and the later to send emails). [...] The name of the contact is usually different for emails (e.g. full name instead of only first name or last name first or use of title in front of name.)''
}
\end{quotation}

\subsection{Activities Performed}

\begin{figure}[htb]
\begin{center}
\includegraphics[width=0.5\textwidth]{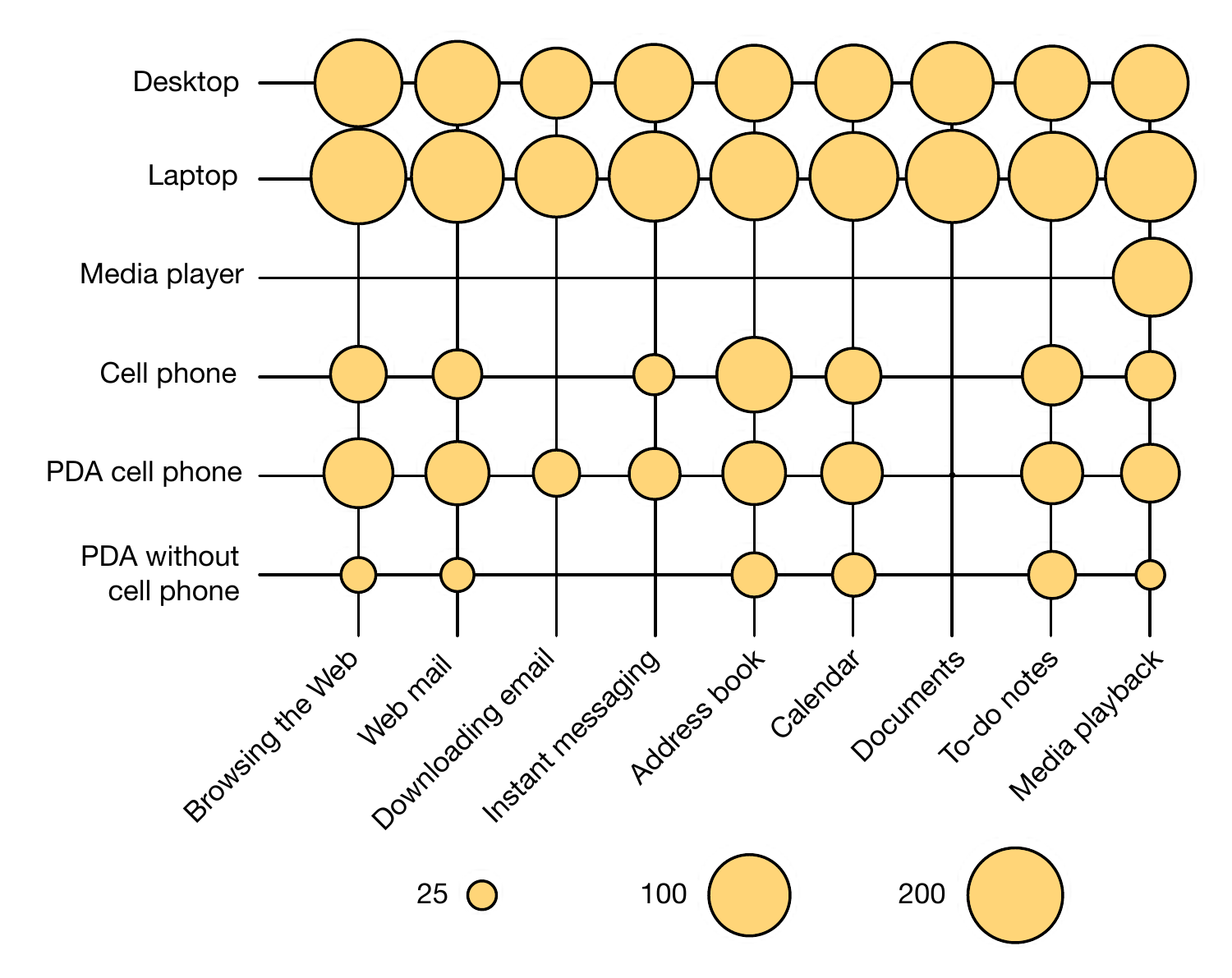}
\caption{Activities performed by users on devices. The diameter of each circle is a logarithmic function of the number of participants who perform a given activity on a given device.}
\label{figure:activities}
\end{center}
\end{figure}

Given the vast array of devices and the features they each support, we wanted to learn what features actually are used. The laptop and desktop were reported as the ones where most computing activities were performed (see Figure \ref{figure:activities}).

Mobile devices such as cell phones and PDAs were used for contact management, making phone calls, and calendaring, and to a lesser extent for browsing and instant messaging. None of the users viewed or edited documents from devices with a smaller form factor than the desktop/laptop.

Users had trouble browsing through their data on small devices, and sometimes even skipped adding more data in order not to ``pollute'' the pool of data already on the device.

\begin{quotation}
\textit{
``I typically will take down someone's email or phone number on a sticky note and then affix it to my cell phone. I find my cell phone's contact navigation to be a real pain. Thus I find it tedious/somewhat-pointless to put more people on there -- after all it will just cause me more pain when I am navigating to people I really want to call.''
}
\end{quotation}

We found several instances of activities performed across devices: users tethered their laptop to their cell phone so that the cell phone's network connection could be used by the laptop, without having to forfeit the richer form factor of the latter. Music was moved off the laptop onto the media player because the media player always was at hand in addition to the laptop.

\section{Conclusion}

As seen from the many examples above, the use of mobile devices involves a lot of compromises and hard decisions that users need to make. There exist concerns stronger than standard metrics such as processor speed, storage and I/O capabilities that propel users to acquire or reject particular devices, including whether the device integrates well into the rest of their information environment, the features it supports, and the other devices that it might replace. Recent advances in mobile technology need to take these into account to design devices and systems that reduce the number of compromises users need to make.


\begin{thebibliography}{10}

\bibitem{abrams_1999_uiml}
M.~Abrams, C.~Phanouriou, A.~Batongbacal, S.~Williams, and J.~Shuster.
\newblock {UIML}: An appliance-independent xml user interface language.
\newblock In {\em Proceedings of the 8th {WWW} conference}, 1999.

\bibitem{bergman_2006_the-project}
O.~Bergman, R.~Beyth-Marom, and R.~Nachmias.
\newblock The project fragmentation problem in personal information management.
\newblock In {\em CHI '06: Proceedings of the SIGCHI conference on Human
  Factors in computing systems}, pages 271--274, New York, NY, USA, 2006. ACM
  Press.

\bibitem{boardman_2003_too-many}
R.~Boardman, R.~Spence, and M.~A. Sasse.
\newblock Too many hierarchies?: The daily struggle for control of the
  workspace.
\newblock In {\em Proc. HCI International 2003}, 2003.

\bibitem{einsenstein_2001_applying}
J.~Einsenstein, J.~Vanderdonckt, and A.~Puerta.
\newblock Applying model-based techniques to the development of uis for mobile
  computers.
\newblock In {\em Proceedings IUI'01: International Conference on Intelligent
  User Interfaces}, pages 69--76. ACM Press, 2001.

\bibitem{florins_2004_graceful}
M.~Florins and J.~Vanderdonckt.
\newblock Graceful degradation of user interfaces as a design method for
  multiplatform systems.
\newblock In {\em IUI '04: Proceedings of the 9th international conference on
  Intelligent user interface}, pages 140--147, New York, NY, USA, 2004. ACM
  Press.

\bibitem{johanson_2001_multibrowsing}
B.~Johanson, S.~Ponnekanti, C.~Sengupta, and A.~Fox.
\newblock Multibrowsing: Moving web content across multiple displays.
\newblock In {\em UbiComp '01: Proceedings of the 3rd international conference
  on Ubiquitous Computing}, pages 346--353, London, UK, 2001. Springer-Verlag.

\bibitem{karger_2006_data}
D.~R. Karger and W.~Jones.
\newblock Data unification in personal information management.
\newblock {\em Commun. ACM}, 49(1):77--82, 2006.

\bibitem{mori_2003_tool}
G.~Mori, F.~Patern\`o, and C.~Santoro.
\newblock Tool support for designing nomadic applications.
\newblock In {\em IUI '03: Proceedings of the 8th international conference on
  Intelligent user interfaces}, pages 141--148, New York, NY, USA, 2003. ACM
  Press.

\bibitem{pyla_2006_multiple}
P.~S. Pyla, M.~Tungare, and M.~P\'erez-Qui\~nones.
\newblock Multiple user interfaces: Why consistency is not everything, and
  seamless task migration is key.
\newblock In {\em Proceedings of the CHI 2006 Workshop on The Many Faces of
  Consistency in Cross-Platform Design.}, 2006.

\bibitem{richter_2005_transformation}
K.~Richter.
\newblock {A Transformation Strategy for Multi-device Menus and Toolbars}.
\newblock In {\em CHI '05: CHI '05 extended abstracts on Human factors in
  computing systems}, pages 1741--1744, New York, NY, USA, 2005. ACM Press.

\bibitem{thevenin_1999_plasticity}
D.~Thevenin and J.~Coutaz.
\newblock Plasticity of user interfaces: Framework and research agenda.
\newblock In {\em Interact}, pages 110--117, Edinburgh, 1999. IFIP.

\bibitem{tungare_2006_personal}
M.~Tungare, P.~S. Pyla, M.~P\'erez-Qui\~nones, and S.~Harrison.
\newblock Personal information ecosystems and implications for design.
\newblock Technical Report cs/0612081, ACM Computing Research Repository, 2006.

\end{thebibliography}
\end{document}